\documentclass[12pt]{article}

\usepackage{xspace}

\newcommand{\ie}{\textit{i.e.}\xspace}
\newcommand{\eg}{\textit{e.g.}\xspace}

\newcommand{\naive}{na\"\i ve\xspace}

\newcommand{\Poincare}{Poincar\'{e}\xspace}

\newcommand{\eq}[1]{(\ref{eq:#1})}  
\newcommand{\Eq}[1]{Eq.~\eq{#1}}         
\newcommand{\Eqs}[1]{Eqs.~\eq{#1}}

\newcommand{\eql}[1]{\label{eq:#1}}

\newcommand{\Ref}[1]{{Ref.~\cite{#1}}}
\newcommand{\Refs}[1]{{Refs.~\cite{#1}}}

\begin{document}

\newcommand{\nc}{\newcommand}

\nc{\beq}{\begin{equation}}
\nc{\eeq}{\end{equation}}
\nc{\beqa}{\begin{eqnarray}}
\nc{\eeqa}{\end{eqnarray}}
\nc{\lsim}{\begin{array}{c}\,\sim\vspace{-21pt}\\< \end{array}}
\nc{\gsim}{\begin{array}{c}\sim\vspace{-21pt}\\> \end{array}}
\nc{\el}{{\cal L}} \nc{\D}{{\cal D}}

\nc{\R}[1]{\ensuremath{R_{#1}}}
\nc{\dimlessR}[1]{\ensuremath{\hat{R}_{#1}}}
\nc{\gs}{\ensuremath{g_{s}}}
\nc{\gYM}{\ensuremath{g_{YM}}}
\nc{\ls}{\ensuremath{l_{s}}}
\nc{\lp}{\ensuremath{l_{P}}}
\nc{\w}{\ensuremath{\omega}}
\nc{\z}{z}

\renewcommand{\thepage}{\arabic{page}}

\begin{titlepage}
\begin{center}
\bigskip

  \hfill YCTP-P11-00\\

  \hfill hep-th/0012033\\

\bigskip
\bigskip
\vskip .5in

{\Large \bf Gravity localization on  string-like defects  in

\smallskip
\smallskip

codimension two and the AdS/CFT correspondence}

\vskip .8 in

{\large Eduardo Pont\'{o}n} and {\large Erich Poppitz }

\vskip 0.2in

\vskip 0.2in {\em
  Department of Physics

  Yale University

  New Haven,
   CT 06520-8120, USA}

\end{center}

\vskip .4in

\begin{abstract}

We study the localization of gravity on string-like defects in
codimension two. We point out that  the gravity-localizing `local
cosmic string' spacetime has an orbifold singularity at the
horizon. The supergravity embedding and the AdS/CFT correspondence
suggest ways to resolve the singularity. We find two resolutions of
the singularity that have a semiclassical gravity description and
study their effect on the low-energy physics on the defect. The
first resolution leads, at long distances, to a codimension one
Randall-Sundrum scenario. In the second case, the infrared physics
is like that of a conventional finite-size Kaluza-Klein
compactification, with no power-law corrections to the
gravitational potential. Similar resolutions apply also in higher
codimension gravity-localizing backgrounds.

\end{abstract}
\end{titlepage}
\setcounter{page}{1}

\baselineskip18pt

\section{Introduction}

The idea that gravity can be localized on a domain wall in a space of
infinite transverse size \cite{RS} has been the subject of much recent
interest, both from phenomenological and theoretical points of view.
In particular, it has been argued that the Randall-Sundrum (RS)
scenario can be given a dual four-dimensional description
\cite{Gubser} via the AdS/CFT correspondence
\cite{MaldacenaConjecture, GKP, Witten}.  In this description, the
four dimensional observable matter is coupled to a four dimensional
``hidden" CFT (strongly coupled, large-$N$) via gravity only (to
define this theory, an ultraviolet cutoff is needed; at energy scales
sufficiently below the cutoff the details of the cutoff are
unimportant).  This duality has been subjected to some quantitative
tests.  Notably, identical power-law corrections to Newton's law are
obtained, which are due to exchange of a continuum of ``Kaluza-Klein"
gravitons in the five dimensional semiclassical gravity description of
RS, or, in the dual CFT description, to (summed-up) hidden sector
loops \cite{Gubser}.

Following the proposal of RS, several generalizations to higher
codimension have been put forward.  Some of these involve localization
to an intersection of domain walls---each a codimension one
object---in some higher dimensional space \cite{kalopernelsonetal}; we
will not study them here.  A different proposal---that gravity can be
localized on a ``stringlike" (codimension two) defect in $AdS_6$ was
made in ref.~\cite{GS} and subsequently generalized to higher
codimension in \cite{GS2}.  It is this proposal we focus on here.

We ask whether there is a dual description like that of the
codimension one RS case and find that the answer is affirmative.
Moreover, as we will see, the spacetime of \cite{GS} has a conical
singularity far from the string.  Semiclassical gravity alone does
not offer guidance towards its resolution.  We will show that the
CFT interpretation of the RS scenario suggests ways to resolve the
singularity, which can have a semiclassical gravity description. We
will investigate how the resolution of the singularity can affect
low-energy quantities on the four dimensional world volume of the
defect, notably the deviation of the gravitational potential from
${1\over r}$.

To better elucidate the previous paragraph, we begin by noting that
the `local cosmic string' metric of \cite{GS} is that of $AdS_6$
with periodic identification of one of the
coordinates:\footnote{More precisely, this is the metric of a
\Poincare patch of AdS. The coordinates we use are related to the
ones of RS \cite{RS} as $\w = R e^{- r/R}$.}
\beq
\label{GSmetric}
d s^2 =   {\w^2\over R^2} \, \left( \eta_{\mu \nu} d x^\mu d x^\nu + R_0^2 d
\theta^2 \right) + R^2 {d \w^2 \over \w^2}~,
\eeq
where $\mu = 0,...,3$ are the Minkowski space coordinates, $\theta$ is
an angular variable, and $R$ is the radius of $AdS_6$.  The ``three
brane" is placed at $\w = R$ and space extends to the $AdS$ horizon
$\w = 0$ \cite{GS}.  The proper size of the space transverse to the brane
is infinite, but its volume is finite and the effective
four-dimensional Planck mass is:
\beq
\label{4dmplanck}
M_{Pl}^2 \sim M_6^4 \int\limits_0^R d \w
\int\limits_0^{2\pi} d\theta \sqrt{g} g^{00} \sim R_0 R M_6^4 ~,
\eeq
where $M_6$ is the six-dimensional Planck mass.  The authors of
\cite{GS} showed that there is a graviton zero mode localized at $\w
=R$ and computed the correction to the gravitational potential in the
four-dimensional theory due to the continuous spectrum of graviton
``Kaluza-Klein" modes:
\beq
\label{newtoncorrectionGS}
V_{(4)}(r) \sim {m_1 m_2 \over M_{Pl}^2} {1\over r} \left(1 +
 {R^3 \over r^3}
\right) ~,
\eeq
where we omitted inessential numerical factors.  The correction to
Newton's law obtained in \cite{GS} has one extra power of $r$ in the
denominator compared to the corresponding correction in the
five-dimensional RS scenario.

The calculation of (\ref{newtoncorrectionGS}) in \cite{GS}
proceeded by imposing particular boundary conditions at the AdS
horizon---the same as in the calculation in the codimension one RS
case \cite{RS, GKR}. There is an important difference between these
two cases, however. The `local cosmic string' spacetime
(\ref{GSmetric}) has a conical singularity at the horizon.  To see
this, note that the metric (\ref{GSmetric}) is obtained from the
$AdS_6$ metric upon identifying one of the Minkowski coordinates
under the action of a discrete translation isometry of Minkowski
space, $\theta \sim \theta + 2 \pi$.  Minkowski space translations,
however, act non-freely on AdS, and it is easy to identify the
fixed points with the \Poincare patch horizon (see \cite{gibbons}
or Appendix A).  Another way the singularity is seen is by noting
that the proper radius of the circle parameterized by $\theta$,
${\cal R}(\w) = \w R_0/R$, shrinks to zero at the horizon.  The
presence of the singularity (even though it is infinitely far away
from the string) and its resolution can significantly affect the
low-energy behavior of the theory on the defect (note that it takes
a finite proper time for geodesics to reach the singularity).

In a string theory framework, as one approaches the horizon,
closed-string winding modes become massless and the geometric
description (\ref{GSmetric}) becomes inadequate.  The advantage of a
string theory embedding is that, as we will see, it offers ways to
deal with the singularity.  In an effective field theory approach, on
the other hand, there is a certain arbitrariness in the boundary
conditions at the singularity, which feeds into the calculation of
(\ref{newtoncorrectionGS}).

To get a guidance as to how the singularity might be resolved and what
the low-energy consequences are, we note first that a dual
interpretation can also be given to a RS setup with gravity localized
on a five dimensional wall in $AdS_6$: five dimensional matter is
coupled to a ``hidden" five dimensional CFT.  This CFT gives rise to
corrections to Newton's law identical to the ones computed from the
classical cutoff-$AdS_6$ gravity.  Wrapping one of the spacelike
directions of the 5d boundary of $AdS_6$ on a circle, as in
eqn.~(\ref{GSmetric}), corresponds to compactifying the dual 5d CFT
(as well as the observable and 5d gravity sectors) on a circle.  This
breaks conformal invariance and induces a nontrivial renormalization
flow of the 5d CFT to a 4d theory.\footnote{The flow of the 5d gravity
to 4d is trivial---for $R_0$ greater than the 5d Planck length gravity
is weakly coupled.  In our discussion we will ignore the observable
sector; we note only that obtaining chiral matter in 4d might require
further orbifolding of the compactified direction.} On general
grounds, depending on the particular CFT and/or the details of the
compactification, there appear to be three possibilities for the end
point of this renormalization flow:
\begin{enumerate}

\item{The 5d CFT on the circle flows to a 4d CFT in the IR.  In this
case, the resulting 4d effective theory is like that of the 4d
Randall-Sundrum case---that of observable matter coupled via
gravity to a ``hidden" CFT.  One expects the same power-law falloff
of the corrections to Newton's law as in the codimension one case.}

{\item The 5d CFT flows to a confining 4d theory, which develops a
mass gap, i.e.  to a trivial infrared fixed point.  In this case, one
expects no power-law corrections to the gravitational potential in 4d.
The resulting description is more like a conventional KK reduction,
with discrete massive graviton modes.}

{\item The 5d CFT flows to a confining 4d theory, which dynamically
breaks some global symmetries.  In the infrared, there are weakly
interacting massless degrees of freedom (goldstone, goldstino,...) in
the hidden sector, giving rise to power-law corrections to the 4d
gravitational potential in the visible sector similar to $1.$ above.}
\end{enumerate}

If a dual gravity description is applicable for any of these possible
end points of the flow, $1.-3.$ should correspond to modifications of
the metric that resolve the singularity at the horizon.  In what
follows, we will show the existence of semiclassical gravity
resolutions of the singularity dual to $1.$ and $2.$ above.  It is not
clear whether (in the deep infrared) $3.$ can have a semiclassical
gravity description---the massless degrees of freedom in the field
theory dual are weakly interacting at low energies and provide a
weakly coupled description of the infrared physics; it is difficult to
contemplate two different weakly coupled dual descriptions of the same
physics.

This paper is organized as follows.  In Sections 2 and 3,
respectively, we consider resolutions of the singularity that
correspond to the flows 1.  and 2.  above.

In Section 2, we follow the supergravity backgrounds corresponding to
the various regions of the flow of the 5d CFT on a circle to a 4d CFT.
We show that the singularity is replaced by a smooth horizon and the
resulting metric, describing the deep infrared region, is nonsingular.
This implies that the infrared physics is like that of the codimension
one RS scenario.  We consider the case of a flow of a 4d CFT on a
circle to a 3d CFT in some detail, since the string theory embedding
and relevant supergravity backgrounds are somewhat simpler, and then
generalize to the 5d $\rightarrow$ 4d flow.

In Section 3, we consider a resolution of the singularity, which, in
the dual field theory, corresponds to imposing antiperiodic boundary
conditions on the fermions of the 5d CFT on the circle.  The resulting
theory flows to a 4d theory with a mass gap.  We note that the use of
string theory dualities in this particular resolution of the
singularity, while suggestive, is not really needed (the resolution
can be simply described in the framework of semiclassical gravity).
We show that the resulting space is nowhere singular and that the
long-distance physics on the defect is like that of conventional KK
compactifications---there are only discrete massive graviton KK modes
and no power-law corrections to Newton's law.

We conclude in Section 4.  For completeness, we give various
technical details, many of which can be found elsewhere in the
literature \cite{gibbons, GK}, in the appendices.  In Appendix A,
we show that the Minkowski translation isometries act non-freely on
AdS, hence identifying the space under the action of a discrete
translation leads to orbifold singularities \cite{gibbons}.  In
Appendix B, we consider the effect of the boundary conditions at
the horizon on the low-energy physics on the defect, from an
effective field theory point of view.  Finally, in Appendix C, we
derive the relation, used in Section 2, between the Neumann Green
function, needed to compute the corrections to the gravitational
potential in the RS scenario, and the Dirichlet kernel, used to
compute correlators in the AdS/CFT correspondence.

\section{Resolution of the singularity 1: flow to a 4d CFT}

As suggested in \cite{Gubser} and further elaborated in \cite{GKR,
GK,Complementarity, SS}, a 4d dual description can be given to the
RS scenario, in which the gravitational backreaction of a 3-brane
in a 5d spacetime of constant negative curvature induces the
localization of gravity near the brane.  The effect of the
noncompact bulk, \ie of the continuum of ``Kaluza-Klein" modes on
the 3-brane physics, can be reproduced in a purely 4d language by
the presence of a (strongly coupled) ``hidden" CFT, which couples
to the 3-brane matter only gravitationally.

This picture is suggested by a generalization of the AdS/CFT
correspondence.  In the metric background
\beq
\label{RSmetric}
ds^{2} =
\frac{\w^{2}}{R^{2}} \left( -dt^{2} + \sum_{i=1}^{3}x^{i}x^{i} \right) +
R^{2}\,\frac{d\w^{2}}{\w^{2}}
\eeq
the 3-brane located at $\w = R$ cuts off the region of AdS space
with $\w > R$.  When one moves the 3-brane to infinity, the AdS/CFT
correspondence relates the string theory partition function on the
background (\ref{RSmetric}) to the partition function of a
4-dimensional CFT ``living" on the boundary of AdS
\cite{MaldacenaConjecture, GKP, Witten}.  Keeping the 3-brane at a
finite $\w$ is interpreted as imposing an UV cutoff on the CFT.  An
important consequence of having a finite cutoff is that the
4-dimensional theory includes gravity (which decouples when the
cutoff is removed).  Thus one can calculate the effects induced by
loops of the cutoff CFT on the gravitational potential produced by
a source. The authors of \Ref{Complementarity} have shown that
these loops exactly\footnote{Due to a nonrenormalization theorem,
see \cite{GKlebanov}, the one loop result for the two-point
function of the stress energy tensor of the  $N = 4$ SYM theory is
exact and applies in the strong coupling limit, where the
comparison to the semiclassical gravity calculation is
appropriate.} reproduce the results of
\Ref{RS}.  In particular, the power-law falloff of the correction
simply follows from the scaling of the two-point correlation
function of the CFT stress-energy tensor at large distances (where
the UV cutoff should not matter).

The above AdS/CFT interpretation of the Randall-Sundrum scenario
suggests ways of resolving the conical singularity in
(\ref{GSmetric}).  It is natural to interpret the corrections to
Newton's law (\ref{newtoncorrectionGS}) as arising from loops of a
hidden 5d CFT with one dimension compactified on a circle.  As was
discussed in the Introduction, these corrections depend on the
infrared behavior of this CFT.

In this Section, we will consider the first possible flow of the 5d
CFT on a circle described in the Introduction---that to a 4d CFT.  We
begin by studying first a simpler problem---the flow of a 4d CFT on
a circle to a 3d CFT.  The dual gravity description is that of a
wrapped 3-brane in a 5-dimensional universe (times a compact
manifold).\footnote{Here the corrections to Newton's law cannot really
be considered small since at low energies the wrapped 3-brane looks
2-dimensional and the leading term is logarithmic.  Nevertheless, it
is useful to study first this case since the supergravity duals
describing the flow of the CFT to the infrared can be described rather
explicitly.} All the issues regarding the singularity are the same as
in the case of a wrapped 4-brane; as we will see, the resolution of
the singularity is also very similar.

We begin by considering the type IIB supergravity solution,
corresponding to a stack of $N$ coincident D3-branes wrapped on a
circle of radius $\R0$:
\beq \eql{D3metric}
ds^{2} = H(r)^{-1/2} \left( -dt^{2} + \sum_{i=1}^{2}x^{i}x^{i} +
\R0^{2}\,d\theta^{2} \right) +  H(r)^{1/2} \left( dr^{2} + r^{2}
d\Omega^{2}_{5} \right)
\eeq
where $H(r) = 1 + \frac{\R3^{4}}{r^{4}}$ and $\R3 = (4 \pi \gYM^{2}
N)^{\frac{1}{4}} \ls$ is assumed to be large enough so that the
supergravity approximation can be trusted.

In the near-horizon limit ($\ls \rightarrow 0$, $r/\ls^2$-fixed) we
can neglect the $1$ in $H(r)$; the metric \Eq{D3metric} then reduces
to
\beq \eql{GShmetric}
ds^{2} = \ls^{2} \left[\frac{u^{2}}{\dimlessR3^{2}} \left( -dt^{2} +
\sum_{i=1}^{2}x^{i}x^{i} + \R0^{2}\,d\theta^{2} \right) +
\dimlessR3^{2}\,\frac{du^{2}}{u^{2}} +
\dimlessR3^{2}\,d\Omega^{2}_{5} \right].
\eeq
where \dimlessR3 is the AdS radius in units of the string length \ls\
and following \cite{MaldacenaConjecture} we expressed the metric in
terms of the ``energy'' variable $u = \frac{r}{\ls^{2}}$.

We note that \Eq{GShmetric}, with the $S^5$ integrated out, is the
same as (\ref{GSmetric}) (less one Minkowskian dimension), if we
appropriately restrict the range of $u$.  As in \Ref{GS} we can think
of this metric as describing the solution outside of a wrapped
3-brane---the ``Planck" brane, not to be confused with the stack of
$N$ wrapped D3 branes whose near-horizon limit is
\Eq{GShmetric}---located at $u_{0} = \dimlessR3 \ls^{-1}$.  The metric
\Eq{GShmetric} localizes gravity close to $u_{0}$ in the same manner
as in the original RS scenario.  We are interested in the
gravitational potential due to a source on the brane when probed by
matter living also on the brane, and more specifically in the
corrections to the (in this case, 2-dimensional) Newtonian potential
induced by the presence of the noncompact bulk.  In the spirit of the
AdS/CFT correspondence, we can think of this scenario as being dual to
a 4d theory where the effects of the bulk are replaced by a cutoff
CFT, weakly coupled to gravity \cite{Gubser, GKR,GK,SS}.  Note,
however, that the conformal invariance is broken not only by the
cutoff but also by the fact that one of the dimensions is compactified
on a circle.\footnote{The transformation $(u, x^{i},\theta)
\rightarrow (\lambda^{-1} u,\lambda x^{i}, \lambda \theta)$ fails to
be an isometry of the metric \Eq{GShmetric} because it changes the
range of $\theta$ from $(0, 2 \pi)$ to $(0, 2 \pi \lambda)$.} We want
to show that the 4d theory flows to a nontrivial infrared fixed point,
where the corrections to Newton's law can be easily estimated.

In order to do this recall that the UV/IR correspondence \cite{SW, PP}
relates the low-energy physics in the CFT to the physics of the
``small-$u$'' region of the supergravity theory in the background
\Eq{GShmetric}.  More precisely, the Newtonian potential at $(X,u)$
due to a pointlike source at $(0,u)$ can be obtained from the Green
function $G_{N}(X,0;u)$ with Neumann boundary conditions imposed at
$u$ \cite{GKR}.  Here $X$ is shorthand for $(x^{i},\theta)$.  But as
shown in \Ref{PS}, if $k \ll \dimlessR3^{-2} u$ we have\footnote{The
authors of \Ref{PS} were interested in the two-point correlation
function $A(k^{2}) = \int\!d x e^{i k x} \langle\mathcal{O}(x)
\mathcal{O}(0)\rangle$.  This is related to the Neumann Green function
by $G_{N}(k;u_{0}) = - A(k^{2})^{-1}$ (see Appendix C).  They also
expressed their results in terms of the coordinate $z =
\frac{\dimlessR3^{2}}{u}$.}
\beq
\tilde{G}_{N}(k; u_0) = Z^{2}(u) \,  \tilde{G}_{N}(k; u) \left(1 +
O\left(\frac{\dimlessR3^{2} k}{u}\right)\right)
\eeq
where $k = \left(k^{i}, \frac{n}{\R0}\right)$ is the momentum
conjugate to $X$.  In other words, up to a wavefunction
renormalization $Z^{2}(u)$, the leading $k$ dependence in
$\tilde{G}_{N}(k; u_0)$ (whose spatial Fourier transform yields the
gravitational potential of a static source at $u_0$) comes from the
region $u \gsim \dimlessR3^{2} k$.

On the other hand, from the metric \Eq{GShmetric}, we see that at any
given $u$ the proper radius of the compact dimension is
$\mathcal{R}(u) = \frac{(\ls u)}{\dimlessR3} \R0$ and therefore the
mass of the KK modes is $m_{KK}(u) \sim \frac{1}{\mathcal{R}(u)}$,
while the mass of the winding modes is $m_{w} =
\frac{\mathcal{R}(u)}{\ls^{2}}$.  These two masses become comparable
at $u_{\ast} \equiv \frac{\dimlessR3}{\R0}$.  It follows that when $k
\lsim \frac{u_{\ast}}{\dimlessR3^{2}} = \frac{1}{\dimlessR3 \R0}$, the
supergravity approximation must break down.

In fact, for $u \R0 \ll \dimlessR3$, it is more appropriate to use the
T-dual description in terms of a D2-brane localized on a circle of
radius $\tilde{R}_{0} = \frac{\ls^{2}}{\R0}$.  The following analysis
is very similar to the one presented in \Ref{SixteenQ} except that now
we have one compact dimension.  For the sake of completeness, we
summarize the main steps.  The type IIA supergravity background
corresponding to the T-dual stack of D2-branes localized on a circle
of radius $\tilde{R}_0$ is:
\beq \eql{D2metric}
ds^{2} = H(\bar{r})^{-1/2} \left( -dt^{2} +
\sum_{i=1}^{2}x^{i}x^{i} \right) + H(\bar{r})^{1/2} \left(
d\bar{r}^{2} + \bar{r}^{2} d\Omega^{2}_{6} \right)~,
\eeq
and the dilaton field is given now by:
\beq \eql{D2dilaton}
e^{\Phi(\bar{r})} = \gs H(\bar{r})^{\frac{1}{4}}.
\eeq
To impose the correct periodicity we take, in the near-horizon limit:
\beq \eql{D2harmonic}
H(\bar{r}) =
\sum_{n=-\infty}^{\infty}{\frac{\R2^{5}}{\mid \bar{r} - \bar{r}_{n}\mid^{5}}}
\eeq
where
$\bar{r}_{n}\equiv (x_{3},x_{4}, \ldots, x_{9})
= (0, 0, \ldots,2 \pi n \tilde{R}_{0})$
and $\R2 = (6 \pi^{2} \gs N)^{\frac{1}{5}} \ls$.  We also identify
$r^{2}=\bar{r}^{2} - x_{9}^{2}$ with the coordinate appearing in
\Eq{D3metric}.  Defining as before energy variables by $u =
\frac{r}{\ls^{2}}$, $\bar{u} = \frac{\bar{r}}{\ls^{2}}$ and $u_{9} =
\frac{x_{9}}{\ls^{2}}$, we find after Poisson resummation that:
\beq
H(r) = \frac{\dimlessR3^{4}}{(\ls u)^{4}}
\left\{1 + 2 \sum_{m = 1}^{\infty}(m u \R0)^{2} \cos(m u_{9} \R0)
K_{2}(m u \R0) \right\}
\eeq
where $K_{2}(x)$ is a modified Bessel function of the second kind
and we used $\gYM^{2} = \frac{\R0}{\ls}\gs$ (with $g_s$---the type
IIA string coupling).  This expression shows that in the limit $u
\R0 \gg 1$ we have $H \simeq \frac{\dimlessR3^{4}}{(\ls u)^{4}}$ up
to exponential corrections, and the metric \Eq{D2metric} can be
written as:
\beq \eql{D2metricLargeu}
ds^{2} = \ls^{2} \left[\frac{u^{2}}{\dimlessR3^{2}}
\left( -dt^{2} + \sum_{i=1}^{2}x^{i}x^{i} \right) +
\frac{\dimlessR3^{2}}{u^{2}} \left( du^{2} + u^{2} d\Omega^{2}_{5} +
\frac{1}{\R0^{2}}\,d\theta^{2} \right) \right].
\eeq
We see that in the T-dual description, the proper radius of the
compact dimension parameterized by $x_{9} =\tilde{R}_{0} \,\theta$
shrinks at large $u$ instead---it becomes of order \ls\ when $u \R0
\sim \dimlessR3$.  This agrees with the limit of validity $u R_0 >$
\dimlessR3 found in the wrapped D3-brane background \Eq{GShmetric}.

The D2-brane gravity background breaks down, in its turn, at small
$u$---this reflects the nonconformality of the D2 brane world volume
theory, which becomes strong in the infrared.  To see this, note that
in the opposite limit $u \R0 \ll 1$, we can just keep the $n = 0$ term
in \Eq{D2harmonic}.  One finds that $H \simeq
\frac{\dimlessR2^{5}}{(\ls \bar{u})^{5}}$ and the effective string
coupling $e^{\Phi}$ becomes of order one at $u \R0 \sim \frac{(6
\pi^{2})^{1/5}}{N^{4/5}} (\gYM^{2} N)$.  In this energy regime, the
supergravity dual is eleven dimensional---we can uplift the
10-dimensional D2-brane background to a solution of eleven dimensional
supergravity; for details see \cite{SixteenQ}.  This solution is, in
its turn, the limit of the M2-brane background when the distances
involved are much larger than $\R{11} = \gs \ls$.  Since we are
interested in the deep infrared, we will skip the uplifted D2 brane
dual and go directly to the M2-brane description.

The near-horizon metric  of a stack of $N$ M2-branes is given by:
\beq \eql{M2metric}
ds^{2} = f(\tilde{r})^{-2/3} \left( -dt^{2} +
\sum_{i=1}^{2}x^{i}x^{i} \right) + f(\tilde{r})^{1/3} \left(
d\tilde{r}^{2} + \tilde{r}^{2} d\Omega^{2}_{7} \right)~,
\eeq
\beq \eql{M2harmonic}
f(\tilde{r}) =
\sum_{n,m=-\infty}^{\infty}{\frac{R^{6}}{\mid \tilde{r} -
\tilde{r}_{n,m}\mid^{6}}}~,
\eeq
where $\tilde{r}_{n,m}\equiv (x_{3}, \ldots,x_{9}, x_{10}) = (0,
\ldots,2 \pi n \tilde{R}_{0},2 \pi m \R{11})$ and $R = (32 \pi^{2}
N)^{\frac{1}{6}} \lp$ (with $\lp$ the 11d Planck length).  Now we
identify $\bar{r}^{2}=\tilde{r}^{2} - x_{10}^{2}$ with the
coordinate in \Eq{D2metric}. This metric has various limits
depending on the relative size of $\R{11}$ and $\tilde{R}_{0}$.
However, we are interested in the deep infrared dynamics of the
CFT, which is mapped to the region $r \ll
\R{11}, \tilde{R}_{0}$.  In this regime, the harmonic function
\Eq{M2harmonic} becomes $f(\tilde{r}) \simeq
\frac{R^{6}}{\tilde{r}^{6}}$ and the metric \Eq{M2metric} can be
written as
\beq \eql{AdS4}
ds^{2} =  \frac{\w^{2}}{R^{2}}
\left( -dt^{2} + \sum_{i=1}^{2}x^{i}x^{i} \right) +
R^{2}\,\frac{d\w^{2}}{\w^{2}} + R^{2}\,d\Omega^{2}_{7} ~,
\eeq
where we defined the new variable $\w = \frac{\tilde{r}^{2}}{R}$.  This
metric describes just $AdS_{4}\times S_{7}$ which shows that there are
no further singularities.  This supergravity background is conjectured
to be dual to a 3d CFT.

To summarize, the previous chain of arguments describes---via the
AdS/CFT correspondence---the flow of a 4d theory which is
approximately conformal in the ultraviolet (where one can ignore the
fact that one dimension is compactified on a circle), to a 3d CFT in
the infrared.  If one interprets the corrections to Newton's law as
arising from loops of this theory, it is appropriate to calculate the
effects of the infrared CFT.  The leading correction comes from the
two-point correlation function of the CFT stress-energy tensor.  But
for a 3-dimensional conformal theory
\beq
\langle T(x) T(0) \rangle \sim \frac{c}{x^{6}}~,
\eeq
where $c \sim (R/l_P)^9 \sim N^{3 \over 2}$ is the central charge
\cite{GKlebanov}, which induces a correction to the Newtonian
potential of order
\beq
\label{deltav}
\Delta V(r) = {1\over M_3} \int d^{2}\!p \, e^{{-i p x}}
\frac{1}{p^{2}}\langle T(p) T(-p) \rangle \frac{1}{p^{2}}
\sim {c \over M_3} \int d^{2}\!p \,\frac{e^{{-i p x}}}{p}
\sim{c \over M_3}\, \frac{1}{r}~,
\eeq
where $M_3$ is the Planck scale of the 3d gravity theory.  We see
that the stringy resolution of the conical singularity has a
dramatic effect on the power-law corrections to the gravitational
potential since a
\naive effective field theory approach would suggest corrections of
order $r^{-2}$, similar to (\ref{newtoncorrectionGS}).

Let us now describe the corresponding construction for the wrapped
4-brane case.  The considerations here closely parallel the ones for
the wrapped 3-brane and we will be correspondingly brief.

The string background whose near-horizon limit is a 6d AdS space
(times a compact space) is that of the type $I^\prime$ D4-D8 brane
system \cite{Oz}.  This background is dual---in the large-$N$ limit,
where $N$ is the number of D4 branes---to a strongly coupled five
dimensional supersymmetric CFT with an $E_{N_f + 1}$ global symmetry
($N_f < 8$ is the number of D8 branes at the O8 plane; for details,
see \cite{Seiberg5d, Oz}).  The gravity background is a fibration of
$AdS_6$ over $S^4$ and is a solution of massive type IIA supergravity
\cite{Oz, cvetic}.

The $AdS_6 \times S^4$ gravity background, therefore, provides a
starting point to study the string embedding of (\ref{GSmetric}).
The metric (\ref{GSmetric})  thus represents the (wrapped) AdS part
of the D4-D8 near-horizon geometry, with one world-volume direction
wrapped on a circle.  The restriction of the radial AdS coordinate
to $\w < R$ is again, in the spirit of the UV/IR correspondence,
interpreted as imposing a cutoff on the dual 5d CFT. The 5d
Newtonian potential, corrected by loops from this ``hidden" CFT,
can be easily evaluated and (ommitting numerical factors) yields,
in the unwrapped ($R_0 \rightarrow \infty$) limit:
\beq
\label{5dnewton}
V_{(5)}(r) \sim {m_1 m_2 \over M_{5}^3} {1\over r^2} \left(1 +
 {R^3 \over r^3}
\right) ~,
\eeq
where $M_5^3 \sim M_6^4 R$ is the 5d Planck scale. The power-law
falloff of  the correction with $r$ can be obtained by scaling from
the two-point function of the energy momentum tensor of the 5d CFT,
as in (\ref{deltav}) above. We conclude that the scaling of the
correction in eqn.~(\ref{newtoncorrectionGS})  is  appropriate at
distances $r \ll R_0$, where the breaking of conformal invariance
is inessential.

As in our discussion of the 4d $\rightarrow $ 3d flow above,
wrapping one world volume direction on the circle breaks conformal
invariance and induces a nontrivial renormalization group flow to a
4d theory. As in going from
\Eq{GShmetric}, via \Eqs{D2metric} and
\eq{M2metric}, to \Eq{AdS4} above, we can use T-duality to study the
dual gravity description of the flow of the compactified CFT to the
infrared.  T-duality along the wrapped worldvolume direction maps the
D4-D8 brane system to the D3-D7 system on a transverse circle of
radius $\tilde{R}_0 = l_s^2/R_0$.  At energies below $1/R_0$ in the
CFT (corresponding to radial distances $\ll \tilde{R}_0$) the
$\tilde{R}_0$ circle is irrelevant and the geometry is approximately
that of the near horizon limit of the D3-D7 gravity background, which
has been studied in \cite{FS, AFM}.  The deep infrared metric
background is $AdS_5 \times \tilde{S}^5$ (the space $\tilde{S}^5$ can
be described as an $S^5$, but with unusual periodicity of one of the
angular variables; for details, see \cite{FS, AFM}).  This shows that
the singularity is resolved and is replaced by a smooth horizon as in
\Eq{AdS4}, as appropriate in the dual description of a theory flowing
to an IR fixed point.  Since, by the UV/IR correspondence, the
nonsingular near-horizon region is the one relevant for the deep
infrared physics, we expect that the infrared physics on the wrapped
4-brane of \cite{GS}---with this particular resolution of the
singularity---is like that of the codimension one RS scenario.

\section{Resolution of the singularity 2: flow to a 4d theory with mass gap}

Another way to resolve the singularity is to modify the metric in the
interior, away from the ``brane" at $\w = R$, in such a way that the
conical singularity of (\ref{GSmetric}) is hidden behind a smooth
horizon.\footnote{This line of thought was suggested to us by S.
Trivedi.} In the dual CFT this modification of the metric has the
interpretation of imposing supersymmetry breaking (antiperiodic on the
fermions) boundary conditions on the compactified circle
\cite{WittenThermalAdS}.  The dual 5d CFT then flows to a
nonsupersymmetric pure Yang-Mills theory in 4d.  The latter has a mass
gap and one does not expect long-range (power-law) deviations from
Newton's law.  Thus, one expects that this type of localization of
gravity is rather similar to the usual KK reduction---there are only
discrete KK modes and no power-law corrections to the gravitational
potential.  This resolution of the singularity realizes the
possibility 2.  pointed out in the Introduction.  In this Section, we
consider the resolution of the singularity by smoothing out the metric
in the interior and show how the above expectations for the infrared
physics are borne out on the gravity side.

We begin by considering the most general solution of the vacuum
Einstein equations in $AdS_6$ with an $SO(2)\times ISO(1,3)$ isometry
\cite{CP, CN, Luty}:
\beq
\label{mostgeneral}
ds^2 = {\w^2 \over R^2}\, \eta_{\mu \nu} d x^\mu d x^\nu +
R^2 {d \w^2 \over \w^2 \left( 1 - {b^5 \over \w^5} \right)} +
{\w^2 \over R^2}\, \left( 1 - {b^5 \over \w^5} \right) R_0^2 d
\theta^2 ~,
\eeq
where $b$ is a yet to be determined constant of integration and
$\theta$ is periodic with period $2 \pi$.  For $b=0$ this is just the
metric (\ref{GSmetric}).  It is easy to see that the metric
(\ref{mostgeneral}) admits half the Killing spinors of $AdS_6$ for $b
= 0$ \cite{LPT} and no Killing spinors at all if $b\ne 0$.  The
analytic continuation ($\theta \rightarrow i t/R_0$, $x^0
\rightarrow i x^4$) of the metric (\ref{mostgeneral}) can be
obtained by a scaling limit from the $AdS_6$ Schwarzschild black
hole solution \cite{BC, cvetic}; the dual CFT interpretation of
this metric background mentioned above is a direct consequence of
this \cite{WittenThermalAdS}.

If $b \ll R$, for $b \ll \w \le R$ the metric approximates that of
\cite{GS}.  On the other hand, at $\w \sim b$ the spacetime is
significantly changed; in particular, for general values of $b$, the
metric has a conical singularity at $\w = b$.  This singularity can be
avoided if one makes a particular choice of $b$.  By considering
(\ref{mostgeneral}) near $\w = b$, it is easily seen that the deficit
angle singularity is absent for
\beq
\label{bvalue}
b = {2 \over 5} \, {R^2 \over R_0}~,
\eeq
and the metric of the 2d transverse space parameterized by ($\w,
\theta$) becomes near $b$ that of the plane in polar coordinates.  The
point $\w = b$ is then a nonsingular horizon.  The singularity of
(\ref{mostgeneral}) at $\w = 0$ is thus hidden ``behind" the horizon;
we should note that this language may be a bit misleading: for $b$
given by (\ref{bvalue}) and $b \le \w \le R$ the space
(\ref{mostgeneral}) is complete and nonsingular---there is no region
``behind" the horizon $\w = b$.  From now on we will consider the
particular value (\ref{bvalue}) of $b$ and discuss the implications
for the ``localization" of gravity and the corrections to Newton's law
on the $\w = R$ brane.\footnote{ We note that for $b$ given by
(\ref{bvalue}), the relation between the 4d and 6d Planck scales is:
$M_4^2 \sim M_6^4 \int \sqrt{g} g^{00} \sim M_6^4 R R_0 \left[ 1 -
\left({2 R\over 5 R_0}\right)^3\right]~\simeq M_6^4 R R_0 $ and it
reproduces (\ref{4dmplanck}) for $b \ll R$.}

In fact, most of the relevant analysis already exists in the
literature: an analogous construction, using compactification of the
$M5$ brane theory on $S^1\times S^1$ with supersymmetry breaking
boundary conditions on one of the $S^1$, was used to study glueball
masses in $QCD_4$ via a scaling limit of the $AdS_7$ black hole
\cite{WittenThermalAdS, OoguriTerning}.  The only modification here is
that by considering only $\w < R$ we cutoff the boundary region of
$AdS$; correspondingly, as in the RS scenario, we need to consider the
non-normalizable modes as well.

To this end, consider the massless scalar wave equation for
$\phi(k,n;\w)$ in the background (\ref{mostgeneral}):
\beq
\label{masslessscalarhorizon}
{1\over R^2\, \w^2}\,  \partial_\w \left[ \w^6 \left(1 - {b^5\over
\w^5}\right)
\partial_\w
\phi(k,n;\w) \right]  - {\w^5 \over \w^5 - b^5} {n^2 R^2 \over R_0^2}
\, \phi(k,n;\w) - k^2 R^2\,
\phi(k,n;\w) = 0 ~,
\eeq
where we have Fourier transformed the field $\phi$ with respect to the
4d coordinates and $\theta$.  The boundary condition at $\w = b$ plays
a crucial role in determining the allowed $k^2$ for the solutions of
this equation.  To determine the boundary condition, note that near
$\w = b$ the Laplacian in (\ref{masslessscalarhorizon}) becomes the
Laplace operator, $\nabla^2$, on the plane (with radial coordinate
$\rho^2 = \frac{4}{5} R^2 (\frac{\w}{b} - 1)$) and the
operator acting on $\phi$ is $\sim \nabla^2 - \kappa^2$, with
$\kappa^2 \sim k^2$.  The solutions of this equation are $J_0(\kappa
\rho)$ and $N_0(\kappa \rho)$.  The $N_0$ solution must be discarded,
as it yields a delta-function singularity ($N_0(x) \sim \log x$
near $x = 0$) when acted upon with $\nabla^2$ and hence does not
obey (\ref{masslessscalarhorizon}) near $\w = b$.  Keeping only the
$J_0$ solution is equivalent to imposing $\partial_\w
\phi(k,n;\w=b)
= 0$.

The values of $k^2$ for which the solutions of
(\ref{masslessscalarhorizon}) obey the Neumann boundary condition
at $\w = b$ determine the $k^2$-plane poles of the
boundary-to-boundary Neumann Green function
$\tilde{G}_N(k;\w,\w^\prime)\vert_{\w=\w^\prime = R}$. These poles,
in turn, determine the masses of the 4d excitations propagating on
the defect.  For $k^2 = 0, n = 0$, the solution is clearly $\phi =
const$.  This is nonnormalizable in the infinite $AdS$ case, but is
normalizable in the $\w < R$ slice.  The contribution of the
constant solution gives rise to the massless graviton and to
Newton's law in the four dimensional theory on the $\w = R$
boundary. Since, as in \cite{WittenThermalAdS, OoguriTerning},
there is no continuum of allowed values of $k^2$ (the quantization
occurs because of the Neumann boundary condition at $\w = b$) we do
not expect the nonanalytic behavior---a logarithmic cut starting at
$k^2 = 0$ in $\tilde{G}_N(k;R,R)$---that leads to power-law
correction in the RS case.  One expects then that the leading
correction is exponential, due to exchange of heavy KK states.

\section{Concluding remarks}

We considered the localization of gravity in codimension two, on a
``stringlike defect" in six dimensions \cite{GS}.  We showed,
guided by the AdS/CFT correspondence, that the resolution of the
singularity of the metric at the horizon can affect the low-energy
physics on the defect and change drastically the long-distance
corrections to Newton's law.  This has a natural interpretation in
the dual CFT description---the long distance corrections to the
gravitational potential due to hidden CFT loops clearly depend on
the infrared physics of the CFT, since compactifying on the circle
breaks conformal invariance and induces a nontrivial flow.  We
enumerated various possibilities and considered two examples of
resolutions of the singularity that have a semiclassical gravity
description.

Singularities like the one considered in this paper will occur also in
generalizations to higher codimension \cite{GS2}.  In particular, one
can wrap $d-4$ of the spatial Minkowski coordinates of the \Poincare
patch of $AdS_{d+1}$ on a Ricci flat compact manifold.  The
corresponding generalization of (\ref{GSmetric}) is still a solution
of the vacuum $AdS_{d+1}$ Einstein equations and leads to localization
of gravity.  As one approaches the horizon, the size of the compact
manifold shrinks, invalidating thus the gravity description.  In a
dual CFT language, similar to the case considered here, a nontrivial
renormalization flow of the CFT is induced.  One expects that in each
case the nature of the infrared dynamics---depending on the details of
the compactification---will influence the long-distance physics on the
defect.

\section{Acknowledgments}

We thank T.  Gherghetta, E.  Katz, and S.  Trivedi for useful
discussions and suggestions.  We are  grateful to the Aspen Center
for Physics for hospitality during the initial stage of this work.
We also acknowledge support of DOE contract DE-FG02-92ER-40704.

\section{Appendix A:}

Here we show that the group of spatial Minkowski translations acts
non-freely on $AdS$.  Therefore, identifying $AdS$ points related
by the action of these isometries leads to singularities at the
fixed points.

 To see this, recall that $AdS_{p+2}$ of unit radius
is defined as the hyperboloid:
\beq
\label{ads}
X_0^2 + X_{p+2}^2 - \sum\limits_{i = 1}^{p+1} X_i^2 = 1~,
\eeq
embedded in $p+3$ dimensional flat space with signature
$(-,+,...,+,-)$.  The \Poincare coordinates $\w, t$, $\vec{x}$, with
$\vec{x}= (x_1,..., x_p)$ and $0 < \w < \infty$, cover half the
hyperboloid:
\beqa
\label{poincarecoord}
X_0 - X_{p+1} &=&  \w  \nonumber
\\ X_0 + X_{p+1} &=& {1\over \w}  + \vec{x}^2 \, \w
-  t^2\, \w
\\ X_i &=&  x_i \, \w~, ~~ i = 1,...,p, \nonumber
\\ X_{p + 2} &=&  t \, \w ~~. \nonumber
\eeqa
{}From (\ref{poincarecoord}) it is easy to see that a Minkowski
spatial translation $\vec{x} \rightarrow \vec{x} + \vec{a}$ acts on
the hyperboloid as:
\beqa
\label{translation}
X_0 - X_{p+1} &\rightarrow&  X_0 - X_{p+1} \nonumber \\ X_0 +
X_{p+1} &\rightarrow&  X_0 + X_{p+1} + \vec{a}^{ 2} (X_0 - X_{p+1})
+ 2
\vec{a} \cdot \vec{X}
\\ X_i &\rightarrow&  X_i + a_i (X_0 - X_{p+1})~, ~ i = 1,...,p, \nonumber
\\ X_{p + 2} &\rightarrow&  X_{p+2}~~. \nonumber
\eeqa
Therefore, translations leave the following points on the hyperboloid
invariant:
\beq
X_{p+1} = X_0~,~~ \vec{a}\cdot\vec{X} = 0~.
\eeq
Using the map (\ref{poincarecoord}), the fixed points are easily
identified with the \Poincare horizon $\w = 0$.  Thus identifying
$AdS_{p+2}$ under the action of a discrete translation leads to
conical singularities at the horizon.

\section{Appendix B:}

{}From an effective 6d gravity point of view the result that the
long distance correction to Newton's law is determined primarily by
the near-horizon geometry, \ie  by the resolution of the
singularity, is somewhat puzzling.  The point is that the
calculation of the potential in the wrapped $AdS_6$ geometry of
ref.~\cite{GS} appears to be insensitive to the
singularity---recall that the singularity is of the orbifold type,
i.e.  no curvature invariants blow up as one approaches the
horizon.  In this Appendix, we study this issue and point out that
the leading correction to the gravitational potential on the defect
depends on the boundary conditions imposed ``at the singularity."
{}From a low-energy point of view, there appears to be an
arbitrariness in the choice of boundary conditions, reflecting the
ignorance of the \naive low-energy theory on the mechanism
resolving the singularity.

The calculation of the correction to Newton's law can be performed by
finding the graviton boundary-to-boundary Green function in
$AdS_{d+1}$ obeying certain boundary conditions; this is equivalent to
the calculations of \cite{RS} (and of \cite{GS} for the wrapped case)
done by decomposing into graviton ``KK" modes.  Since the appropriate
graviton propagator can be expressed in terms of the massless scalar
propagator \cite{GKR}, for our purpose it will be sufficient to study
the scalar Green function.

We begin by studying the massless scalar Green function in the wrapped
$AdS_{d+1}$ background with particular attention to the boundary
conditions.  We use the metric\footnote{This metric is related to the
metric in (\ref{GSmetric}) by the change of variables $\z = {R\over
\w}$.  This parametrization is more convenient for the purpose of
deriving the propagator.}
\beq
\label{PoincareMetric}
d s^2 =
{1\over \z^2}\eta_{\mu \nu} d x^\mu d x^\nu + R^2 {d \z^2 \over \z^2}~,
\eeq
where $\eta_{\mu \nu}$ is the $d$ dimensional Minkowski metric (mostly
plus). To study the wrapped case, we will imagine that one of
the $d-1$ spatial coordinates is compactified on a circle of
coordinate radius $R_0$, as in (\ref{GSmetric}).  We will let the
coordinate $\z$ change between $\z = \epsilon$ and $\z =\infty$ (the AdS
horizon in these coordinates); thus the ``brane" is located at $\z =
\epsilon$.  We denote the Fourier transform of the massless scalar
field with respect to the Minkowski coordinates by $\phi(k, \z)$, where
$k$ is the $d$-dimensional momentum vector (with a discrete component
$n/R_0$ when a direction is compactified).  The scalar Laplace
operator ($\nabla^2 \equiv g^{-1/2} \partial_M g^{MN} g^{1/2}
\partial_N$; $g^{1/2} \equiv R \z^{-d-1}$) thus becomes:
\beq
\label{scalarlaplace}
\nabla^2_\z \phi(k, \z) = {\z^2 \over R^2} \phi^{\prime\prime}(k, \z) -
{(d-1) \z \over R^2} \phi^{\prime}(k,\z) - k^2 \z^2 \phi(k, \z) ~,
\eeq
where primes denote derivatives with respect to the $\z$ coordinate.
For any two fields $\phi_1(k,\z)$ and $\phi_2(k,\z)$ we can, upon using
(\ref{scalarlaplace}) and integrating by parts, obtain the Green
formula:
\beqa
\label{greenformula}
\int\limits_{\epsilon}^\infty d \z \sqrt{g} \left[ \phi_2(k,\z)
\nabla^2_\z \phi_1(k,\z) - \phi_2(k,\z) \nabla^2_\z \phi_1(k,\z) \right]
\nonumber \\
\hspace{1cm}
= \sqrt{g}\,g^{\z\z} \left[\phi_2(k,\z) \partial_\z \phi_1(k,\z) -
\phi_2(k,\z) \partial_\z \phi_1(k,\z)\right]\bigg\vert_{\z = \epsilon}~.
\eeqa
If we take now $\phi_1(k,\z) = G(k;\z,\z^\prime)$, obeying
\beq
\label{greenfunctioneqn1}
\sqrt{g}\,\nabla^2_\z G(k;\z,\z^\prime) =
\delta(\z - \z^\prime)~,
\eeq
and $\phi_2(k,\z) = \phi(k,\z)$---a solution of the bulk equation
$\nabla^2_\z \phi(k,\z) = J(k,\z)$, and substitute in
(\ref{greenformula}) we obtain:
\beqa
\label{greenfinal}
\phi(k, \z^\prime) &=& \int\limits_{\epsilon}^\infty d \z \sqrt{g} \,
G(k; \z, \z^\prime) J(k,\z)
\nonumber \\
&+& \sqrt{g(\z)}\,g^{\z\z}(\z) \left[ \phi(k,\z)
\partial_\z G(k; \z, \z^\prime) - G(k; \z, \z^\prime)
\partial_\z \phi(k,\z) \right]\bigg\vert_{\z = \epsilon}~.
\eeqa
The Green formula (\ref{greenfinal}) is important in determining the
consistency of various boundary conditions on $\phi$ and $G$ at $\z =
\epsilon$.

The calculation of the ``Newtonian" potential---here we really are
computing the scalar Green function; for its relation to the
graviton one, see \cite{GKR}---requires finding the Green function
obeying a Neumann (N) boundary condition $\partial_\z
G_N(k;\z,\z^\prime)\big \vert_{\z = \epsilon} = 0$ at $\z =
\epsilon$.  One then computes the boundary-to-boundary Green
function $G_N(k; \epsilon, \epsilon)$ for spacelike momenta (with
$k_0 = 0$).  The potential at the boundary, $\varphi(r, \epsilon)$,
due to a static source $J$ at $\z = \epsilon$ is found from
eqn.~(\ref{greenfinal})\footnote{The Neumann boundary condition at
$\z = \epsilon$
assures that the first boundary term in (\ref{greenfinal}) does not
contribute.  One is then free to specify the normal derivative,
$\partial_\z \phi(k,\z)$, at the boundary.  In the cases of interest
one imposes a $Z_{2}$ reflection symmetry about the brane at $\z =
\epsilon$, so the normal derivative of $\phi(k,\z)$ at the boundary
actually vanishes and the potential is just given by the bulk
integral in (\ref{greenfinal}).} after a $d-1$ dimensional spatial
Fourier transform.  If a direction is wrapped, the Fourier
transform w.r.t. the wrapped component of $k$ is replaced by a
discrete sum.  In order to find the leading long-distance behavior
of $V(r)$, we need the small-$k$ expansion of the
boundary-to-boundary propagator $G_N(k; \epsilon, \epsilon)$.  The
$k^{-2}$ term yields the leading term, giving Newton's law in $d$
dimensions after the Fourier transform. The terms containing
positive integer powers of $k^2$ yield local terms in
$V(r)$---delta function and its derivatives---and thus do not
affect the long-distance behavior.  The leading long-distance
correction to Newton's law arises from the first nonanalytic
term---logarithm of $k^2$ for $d$-even or a fractional power of
$k^2$ for odd $d$.

Even though the horizon at $\z = \infty$ is not a boundary in the
same sense as the hyperplane $\z = \epsilon$, the \Poincare patch
parameterized by Eq.(\ref{PoincareMetric}) can be continued beyond
the horizon and it is necessary to specify appropriate boundary
conditions in order to obtain a unique Green function. We are
interested in exploring the possible low-energy effects
parameterized by this freedom.

To begin, note that the general solutions of (\ref{greenfunctioneqn1})
for $\z \ne \z^\prime$ are:
\beqa
\label{greensN1}
G_<(k;\z,\z^\prime) &=& f_1(k; \z^\prime) \z^{d \over 2} H_{d\over
2}^{(1)}(ikR\z) + f_2(k; \z^\prime) \z^{d\over 2} H_{d\over
2}^{(2)}(ikR\z) ~,~{\rm for} ~ \z < \z^\prime~,
\nonumber \\
G_>(k;\z,\z^\prime) &=& g_1(k; \z^\prime) \z^{d \over 2} H_{d\over
2}^{(1)}(ikR\z) + g_2(k; \z^\prime) \z^{d\over 2} H_{d\over
2}^{(2)}(ikR\z)~,~{\rm for} ~ \z > \z^\prime~,
\eeqa
where $k = \sqrt{k^2}$ for spacelike Minkowski momenta and $k = i
\sqrt{|k^2|}$ for timelike momenta.  We note that since $H^{(2)}(y)
\sim e^{- i y}$ as $y\rightarrow \infty$, the second term in $G_>$
exponentially grows for spacelike momenta as $\z$ approaches the
horizon.  For timelike momenta and positive frequency $k_0$ the
$H^{(2)}$ term in $G_>$ represents a wave moving in   from the
 past
horizon ($\sim e^{i k_0 t + i |k| R \z}$),
  while the first term
  represents a wave traveling towards the future horizon.

In the calculation of ref.~\cite{GKR} the Hartle-Hawking boundary
condition, corresponding to keeping only the wave moving towards
the future horizon---in other words setting $g_2 \equiv 0$ in
(\ref{greensN1})---was imposed.  This is also the natural boundary
condition to impose in D-brane absorption cross section
calculations that lead to the AdS/CFT correspondence \cite{GKP}.

In what follows, we will proceed without imposing the
Hartle-Hawking boundary condition at the horizon.  The rationale is
that the resolution of the singularity at the horizon changes the
``potential" in the near horizon region and induces a ``reflected"
wave; therefore a more general condition, allowing for both
reflected and transmitted waves, should be imposed at large $\z$.

Imposing now the Neumann condition $\partial_\z
G_N(k;\z,\z^\prime)\big\vert_{\z = \epsilon} = 0$ on $G_<$ and the
appropriate discontinuity at $\z = \z^\prime$ to reproduce the delta
function in (\ref{greenfunctioneqn1}), we find that the Neumann
function $G_N$ is given by (\ref{greensN1}), where $g_{1}$ and
$f_{1,2}$ are:
\beqa
\label{greensN2}
g_1(k; \z) &=& { 1 \over W(k;\epsilon)} \left( M_2(k;\z) - M_1(k;\z)
{M_{2,\z}(k;\epsilon) \over M_{1,\z}(k;\epsilon)} \right) -
g_2(k;\z){M_{2,\z}(k;\epsilon) \over M_{1,\z}(k;\epsilon)} ~ ,
\nonumber \\
f_1(k;\z) &=& g_1(k;\z) - {M_2(k;\z) \over W(k;\epsilon)} ~,\\
f_2(k; \z) &=&g_2(k;\z) + {M_1(k;\z)\over W(k;\epsilon) } ~,
\nonumber
\eeqa
where we defined the functions $M_{1,2}(k;\z)$ as
\beq
\label{M12}
M_{1 [2]}(k;\z) \equiv \z^{d\over 2} H^{(1)[(2)]}_{d\over 2}(ikR\z)~,
\eeq
and $M_{1[2],\z}(k;\z) \equiv \partial M_{1[2]}(k;\z)/\partial \z$.  The
function $g_2(k;\z)$ is arbitrary for now; as $g_2 \rightarrow 0$ we
obtain the Hartle-Hawking Green function given in \cite{GKR}.  In
eqn.~(\ref{greensN2}) we have introduced
\beq
\label{wronskian} W(k;\z) \equiv { M_2(k;\z)
M_{1,\z}(k;\z) - M_1(k;\z) M_{2,\z} (k;\z)\over R \z^{d -1} }~,
\eeq
which, being the Wronskian of two solutions of the homogeneous
equation, is constant (this can be inferred from
(\ref{greenformula})) and can be evaluated, for example at the
boundary at $\z = \epsilon$. {}From eqn.~(\ref{greenfinal}), it
follows that for consistency the Green function has to also satisfy
eqn.~(\ref{greenfunctioneqn1}) when the Laplacian acts on the
second argument.  This implies, upon inspection of
(\ref{greensN2}), that the function $g_2(k;\z)$ has to also solve
the homogeneous Laplace equation, \ie
\beq
\label{g2}
g_2(k;\z) = a(k) M_1(k;\z) + b(k) M_2(k;\z)~,
\eeq
where $M_{1,2}$ were defined in (\ref{M12}) and $a(k), b(k)$ are still
arbitrary functions of momentum (we note that for any $g_2(k;\z)$ of
the above form one obtains the correct discontinuity of
$\partial_{\z^\prime} G(k;\z,\z^\prime)$ at $\z = \z^\prime$).  Consistency
of the Green formula (\ref{greenfinal}) requires also that
$\partial_{\z^\prime} G_>(\z, \z^\prime)\vert_{\z^\prime = \epsilon} = 0$
leading to the relation
\beq
\label{abrelation1}
a(k) = - b(k) ~{M_{2,\z} (k; \epsilon) \over M_{1,\z} (k; \epsilon)}~,
\eeq
leaving one arbitrary function of momentum, $b(k)$, in the Green
function. Putting Eqs.(\ref{greensN1}), (\ref{greensN2}),
(\ref{g2}), and (\ref{abrelation1}) together we get the Neumann
Green function
\beqa
\label{greensfinalsolution}
G_{N}(k;\z,\z^\prime) & = & {M_{1}(k;\z_{\scriptscriptstyle{>}}) \over
W(\epsilon)} \left(M_{2}(k;\z_{\scriptscriptstyle{<}}) -
M_{1}(k;\z_{\scriptscriptstyle{<}})
{M_{2,\z}(k;\epsilon) \over M_{1,\z}(k;\epsilon)} \right) \\
& & \mbox{} + b(k) \left(M_{2}(k;\z) - M_{1}(k;\z)
{M_{2,\z}(k;\epsilon)\over M_{1,\z}(k;\epsilon)} \right)
\left(M_{2}(k;\z^{\prime}) - M_{1}(k;\z^{\prime})
{M_{2,\z}(k;\epsilon)\over M_{1,\z}(k;\epsilon)} \right)
\nonumber
\eeqa
where $\z_{\scriptscriptstyle{>}}$ ($\z_{\scriptscriptstyle{<}}$)
is the greatest (smallest) of $\z, \z^{\prime}$. This formula
displays the fact that the general solution can be written as the
sum of the Green function obeying the Hartle-Hawking boundary
condition (given by the first term) plus an arbitrary solution of
the homogeneous equation, given by the second term.

In this language the resolution of the singularity would amount to
the specification of the boundary conditions at the horizon which
would determine $b(k)$.  As an example, consider the case where we
regulate the singularity by ``hiding" it behind a horizon as in
equation (\ref{mostgeneral}).  In the limit $b \rightarrow 0$, we
recover the background we are interested in (where the variables in
Eqs.(\ref{mostgeneral}) and (\ref{PoincareMetric}) are related by
$\z = \frac{R}{\w})$.  However, as we remarked in Section 3, for
arbitrary values of  $b$, the point $\z = \frac{R}{b}$ corresponds
to a conical singularity with deficit angle $\Delta
\theta = 2 \pi (1 - \frac{5}{2} \frac{\R0 b}{R^2})$. The deficit
angle gives rise to a delta function singularity in the Einstein
tensor, which can be interpreted as a brane (see also
\Refs{CP,Luty}). In particular, if the field in question (gravity
or, in this case, the scalar field) couples to this brane there
could be a backreaction from the brane when a source is turned on
elsewhere. One can imagine encoding the backreaction in some
complicated form of boundary conditions at the brane (\eg some
linear combination of Dirichlet and Neumann boundary conditions
with coefficients that could very well depend on the 4-momentum
$k$). This would correspond to the freedom parameterized by $b(k)$
in Eq.(\ref{greensfinalsolution}) and could have important effects
as we saw in Sections 2 and 3.

\section{Appendix C:}

In this Appendix, we derive the relation between the
boundary-to-bulk propagator, relevant for the calculation of the
two-point correlation function in the AdS/CFT correspondence
\cite{GKP,Witten} and the Neumann propagator, relevant to determine
the gravitational potential (see also \Ref{GK}).  As usual, we
restrict ourselves to the case of a scalar field. The calculation
of the two point correlator proceeds by solving the bulk field
equations with a specified value for the field at the boundary, \ie
by imposing Dirichlet (D) boundary conditions.  The propagator
appropriate for this kind of boundary conditions satisfies
$G_{D}(k;\z = \epsilon,\z^\prime) = 0$.  Then Eq.(\ref{greenfinal})
(with $J(k,\z) = 0$) gives the desired solution:
\beq
\label{Dirichletsolution}
\phi(k, \z^\prime) = \sqrt{g(\z)}\,g^{\z\z}(\z) \phi_{0}(k)
\partial_\z G_{D}(k; \z, \z^\prime) \bigg\vert_{\z = \epsilon}~,
\eeq
where $\phi(k, \z^\prime = \epsilon) = \phi_{0}(k)$.  The
boundary-to-bulk propagator is just
\beq
\label{boundarybulk}
K(k;z^{\prime}) = \sqrt{g(\z)}\,g^{\z\z}(\z)
\partial_\z G_{D}(k; \z, \z^\prime) \bigg\vert_{\z = \epsilon}~,
\eeq
and, from Eq.(\ref{Dirichletsolution}) evaluated at $z^{\prime} =
\epsilon$, it satisfies
\beq
K(k;z^{\prime} = \epsilon) = 1.
\eeq
The two-point CFT correlation function can then be obtained via
\cite{GKP, Witten}
\beq
\label{correlator}
A(k^{2}) \equiv \int d x e^{i k x} \langle\mathcal{O}(x)
\mathcal{O}(0)\rangle = \sqrt{g(\z)}\,g^{\z\z}(\z)\partial_\z K(k;z)
\bigg\vert_{\z = \epsilon}~.
\eeq

Now note that the Green formula  Eq.(\ref{greenfinal}) also allows
us to write the same solution for $\phi(k, \z^\prime)$ in terms of
the Neumann propagator if we specify the ``correct'' normal
derivative, $\partial_\z
\phi(k,\z) \vert_{\z = \epsilon}$, at the boundary:
\beq
\label{Neumannsolution}
\phi(k, \z^\prime) = -\sqrt{g(\z)}\,g^{\z\z}(\z) G_{N}(k; \z, \z^\prime)
\partial_\z \phi(k,\z) \bigg\vert_{\z = \epsilon}~
\eeq
where
\beq
\partial_\z \phi(k,\z) \bigg\vert_{\z = \epsilon} =
\phi_{0}(k) \partial_\z K(k;z) \bigg\vert_{\z = \epsilon}
\eeq
is obtained by differentiating Eq.(\ref{Dirichletsolution}) (we
used the definition (\ref{boundarybulk}) of $K$). With this
boundary condition, the solution (\ref{Neumannsolution}) must be
the same as solution (\ref{Dirichletsolution}) (at least in the
Euclidean case). Thus, using the AdS/CFT relation
(\ref{correlator}), we find
\beq
\label{generalrelation}
K(k;z^{\prime}) = -G_{N}(k; \epsilon, \z^\prime) A(k^{2})
\eeq
and setting $z^{\prime} = \epsilon$ we obtain the final relation
\beq
G_{N}(k; \epsilon, \epsilon) = - A(k^{2})^{-1}.
\eeq
Note that this relation remains valid if we replace $\epsilon$ by
an arbitrary $\z$ (also in the definition of $A$,
(\ref{correlator})).

\end{document}